# ThermInfo: Collecting, Retrieving, and Estimating Reliable Thermochemical Data

Ana L. Teixeira,[†,‡,*] Rui C. Santos,[‡] João P. Leal,[‡,§] José A. Martinho Simões,[‡] and Andre O. Falcao,[†]

† LaSIGE, Departamento de Informática, Faculdade de Ciências, Universidade de Lisboa, 1749-016 Lisboa, Portugal; ‡ Centro de Química e Bioquímica, Faculdade de Ciências, Universidade de Lisboa, 1749-016 Lisboa, Portugal;  § Unidade de Ciências Químicas e Radiofarmacêuticas, Instituto Tecnológico e Nuclear, Instituto Superior Técnico, Universidade Técnica de Lisboa, 2686-953 Sacavém, Portugal.

**ABSTRACT**.

Standard enthalpies of formation are used for assessing the efficiency and safety of chemical processes in the chemical industry.  However, the number of compounds for which the enthalpies of formation are available is many orders of magnitude smaller than the number of known compounds.  Thermochemical data prediction methods are therefore clearly needed.  Several commercial and free chemical databases are currently available, the NIST WebBook being the most used free source.

To overcome this problem a cheminformatics system was designed and built with two main objectives in mind: collecting and retrieving critically evaluated thermochemical values, and estimating new data.  In its present version, by using cheminformatics techniques, ThermInfo allows the retrieval of the value of a thermochemical property, such as a gas-phase standard enthalpy of formation, by inputting, for example, the molecular structure or the name of a compound.  The same inputs can also be used to estimate data (presently restricted to non-polycyclic hydrocarbons) by using the Extended Laidler Bond Additivity (ELBA) method. The information system is publicly available at http://www.therminfo.com or http://therminfo.lasige.di.fc.ul.pt. ThermInfo's strength lies in the data quality, availability (free access), search capabilities, and, in particular, prediction ability, based on a user-friendly interface that accepts inputs in several formats.

\* Corresponding author.
*E-mail address:* ateixeira@lasige.di.fc.ul.pt

# INTRODUCTION

Chemical information keeps growing fast [1,2]. According to the Chemical Abstracts Service [3], there are currently more than 66 million known organic and inorganic substances and approximately 12,000 new entries are added daily. Cheminformatics plays, therefore, an increasingly important role, not only in archiving and retrieving information but also in implementing property estimation methods through suitable software.

There is a number of databases covering a variety of chemical and physical properties [4]. Some are hard copies (predominantly the oldest ones), other are electronic, and several have both formats [5-8]. The present trend, however, is creating either public (free access) or commercial web-based information systems [9-23]. The advantages and disadvantages of both options have been discussed by some authors [24-29].

Thermodynamic values are of major importance in chemistry and chemical engineering [30]. However, the main repositories for those values are printed handbooks and literature reviews [31-47] and only a minority of the web-based information systems dedicated to thermodynamic properties provide full[11,16,48-50] or limited[12,16,18,20] free access to their database.

Standard enthalpy of formation values are essential for assessing the efficiency and safety of any chemical process [51]. Enthalpies of formation can be obtained experimentally but this often requires complex and expensive methodologies [52]. *Ab initio* methods can also be reliable tools to determine enthalpies of formation [53,54]. However, the most accurate quantum chemistry methods are still computationally too expensive for large molecules [55]. Moreover, these methods are often validated with small molecules and their validity for large molecules is assumed. To avoid this problem, the theoretical methods are often used together with homodesmotic reactions [56], which are designed to maximize error cancellation.

Despite the work of experimental and theoretical chemists, it is thus not surprising that the number of compounds for which the enthalpies of formation are available is many orders of magnitude smaller than the number of known compounds. Thermochemical data prediction methods are therefore badly needed [24].

One of the most accurate ways to estimate enthalpies of formation relies on additivity methods [31,37,40,57-63], such as the so-called Benson group method (initially proposed in 1958), whose parameters (group enthalpies) have been refined and extended by several authors [57,58]. Another additivity method that is simple to use and very attractive to chemists (since it deals with bond enthalpies), is the Laidler method [63]. In this scheme the main parameters are assigned to the chemical bonds, and therefore should reflect the strengths of those bonds. The Laidler method has been recently refined for a variety of hydrocarbon families [59,60] and this new parameterization, called the Extended Laidler Bond Additivity (ELBA) method, improves the reliability of estimates (even for very large and bulky compounds) and extends the method to new families of hydrocarbon compounds.

As explained elsewhere [60], the prediction of accurate data by the ELBA method demands a large number of empirical parameters. Moreover, all the important parameters, including non-bonded interactions, must be recognized a priori, in order to obtain accurate predictions. The identification of all the parameters is a time-consuming task and often requires some familiarity with the prediction method. These disadvantages can only be removed by implementing an application suitable for non-expert users.

There are several computer and web-based applications based on empirical additivity schemes for estimating thermochemical data of organic compounds. Examples of the former include THERM/EST[64], CHETAH[51], DIPPR 801/DIADEM[65], DETHERM Software Suite[66], and NIST ThermoData Engine[67]. The latter applications include S&P[68], DDBSP-ARTIST[69], and DDBST-UNIFAC[70].

THERM/EST (NIST Estimation of Thermodynamic Properties for Organic Compounds) is a computer program developed to estimate thermodynamic data based on the principles of group additivity developed by Benson et al.[57] and later extended by Domalski et al.[58]. The commercial version of this software was discontinued [64]. Another application is the NIST Structures and Properties program [68], which also features an implementation of Benson's group additivity method [57], using a graphic interface that allows the user to draw the molecule and estimate several thermochemical properties. The commercial version of this software was also discontinued but it is still available at the NIST WebBook site [68]. Another NIST product, the ThermoData Engine software, provides critically evaluated thermodynamic and transport property data [67]. Finally, the ASTM Computer Program for Chemical Thermodynamics and Energy Release Evaluation (CHETAH)[51] software is a commercial tool for predicting thermochemical properties and certain hazards associated with a chemical compound or a reaction. This is accomplished through the knowledge of the molecular structure(s) of the components involved, and once again by an implementation of Benson's group additivity method [57].

Having in mind all the strengths and limitations of both the offline and online applications we set off to develop a free access information system, which we called ThermInfo, with the following features: (1) it should be user friendly; (2) all the data should be critically evaluated, either by ourselves or by a trusted source; (3) the data should include values for a wide range of compounds, viz. long-lived and transient organic, inorganic, and organometallic molecules in the gas- and condensed-phases; (4) a variety of empirical methods, selected on the basis of their reliability to predict data, would be included; (5) the parameterization of those empirical methods should be easy to recalculate on the basis of new data. In addition, we would pursue the search of new estimation procedures, based on structure-energy relationships and machine learning methods.

The initial ThermInfo dataset was based on Pedley's compilations of organic compounds properties [40,41]. The ELBA method (presently restricted to non-polycyclic hydrocarbons) [60], was selected for thermochemical prediction. This method yields reliable estimates but, as stressed above, it is only suited to computational use due to the large number of parameters involved.

The following section of this manuscript describes the database, the data assessment, the method used to predict properties, and the system basic architecture and requirements. Then the database implementation details as well as the technology options to implement the system are explained. After describing ThermInfo's system architecture and its dissemination facilities, the main conclusions and directions for future work are presented.

## CONSTRUCTION AND CONTENT

### Available data and their assessment

As mentioned before, the ThermInfo starting point was the experimental thermochemical database taken from two critically evaluated compilations by Pedley and coworkers,[40,41] together with the associated structural data. Although that thermochemical database is not error-free, it has the important advantage of being thermodynamically consistent, while some others are mere compilations. Detected faults will be addressed in future revisions, although they have a negligible impact on the ELBA parameters (see below).

New databases and individual values will be added in the near future. The procedure will be as follows:

New data may come from other compilations or from the primary literature. In both cases, a critical evaluation of the values will be performed. Checks of the structural information will also be performed, based on different sources. When several values are available for a given property of the same compound, a single value will be chosen to be the "recommended value", although the rest of the values will be stored in the database. This choice will rely on several criteria, such as the experimental method(s) used, and also on consistency tests performed by applying available estimation methods and/or high-level quantum chemistry calculations.

### The ELBA method and its implementation

The ELBA version used in ThermInfo includes a set of 165 parameters [60]. Each one of these new parameters has an assigned physical meaning (i.e. they are not fudge parameters). An example of the method, illustrating the calculation of gas- and liquid-phase standard enthalpies of formation, is presented in Table 1.

**Table 1.** ELBA parameters required to estimate the gas- and liquid-phase standard enthalpies of formation at T=298.15 K (kJ mol$^{-1}$) of bicyclohexyl and 4-methyl-1-*tert*-butylbenzene.

| Compound name | bicyclohexyl | 4-methyl-1-*tert*-butylbenzene |
|---|---|---|
| SMILES | C1(C2CCCCC2)CCCCC1 | CC(C1=CC=C(C)=C1)(C)C |
| Compound structure | 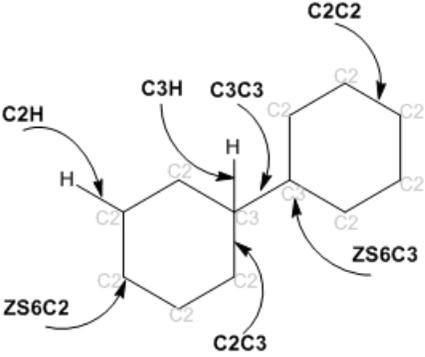 | 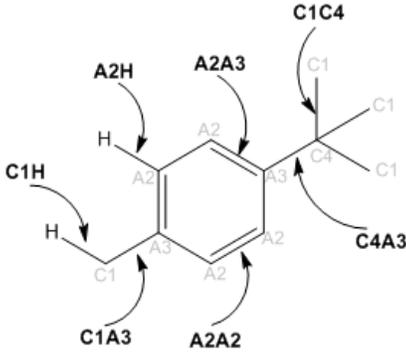 |
| Sum of ELBA parameters | 8 C2C2 + 4 C2C3 + 1 C3C3 + 20 C2H + 2 C3H + 10 ZS6C2 + 2 ZS6C3 | 3 C1C4 + 12 C1H + 1 C1A3 + 1 C4A3 + 2 A2A2 + 4 A2A3 + 4 A2H |

| $\Delta_f H_m^\circ$ (g) (kJ mol$^{-1}$) | | | | |
|---|---|---|---|---|
| | Experimental | 215.7±1.5$^a$ | Experimental | 57.6±1.0$^b$ |
| | Estimated | 214.6 | Estimated | 56.0 |
| $\Delta_f H_m^\circ$ (l) (kJ mol$^{-1}$) | Experimental | 273.7±1.4$^a$ | Experimental | 109.7±0.9$^b$ |
| | Estimated | 274.0 | Estimated | 108.4 |

$^a$ Reference [40];  $^b$ Verevkin, S. P. Thermochemical properties of branched alkylsubstituted benzenes. *J. Chem. Thermodyn.* **1998**, *30*, 1029–1040.

The automatic generation of structural descriptors from a computer-readable representation of a chemical structure diagram was not a straightforward task. These representations are analyzed automatically to derive the frequencies of occurrence of the ELBA parameters using a procedure that iterates the molecular structure (bonds and atoms) and extract the 165 ELBA parameters based on a set of structural characteristics (viz. the number of atoms, number of bonds, number of single, double, triple and aromatic bonds, number of rings, atom multiplicity, bond order, atom environment, maximum and minimum bond order, ring size, and cis/trans configurations) and a set of interrelated rules (Figure 1). The structure of the compounds is encoded in the widely used notations in chemical information systems, the SMILES and 3D MDL MOL File [71-73]. To validate ELBA implementation, along with other extensive tests, enthalpies of formation were manually calculated by the ELBA method for a set of more than 450 compounds and compared with the data computed by ThermInfo.

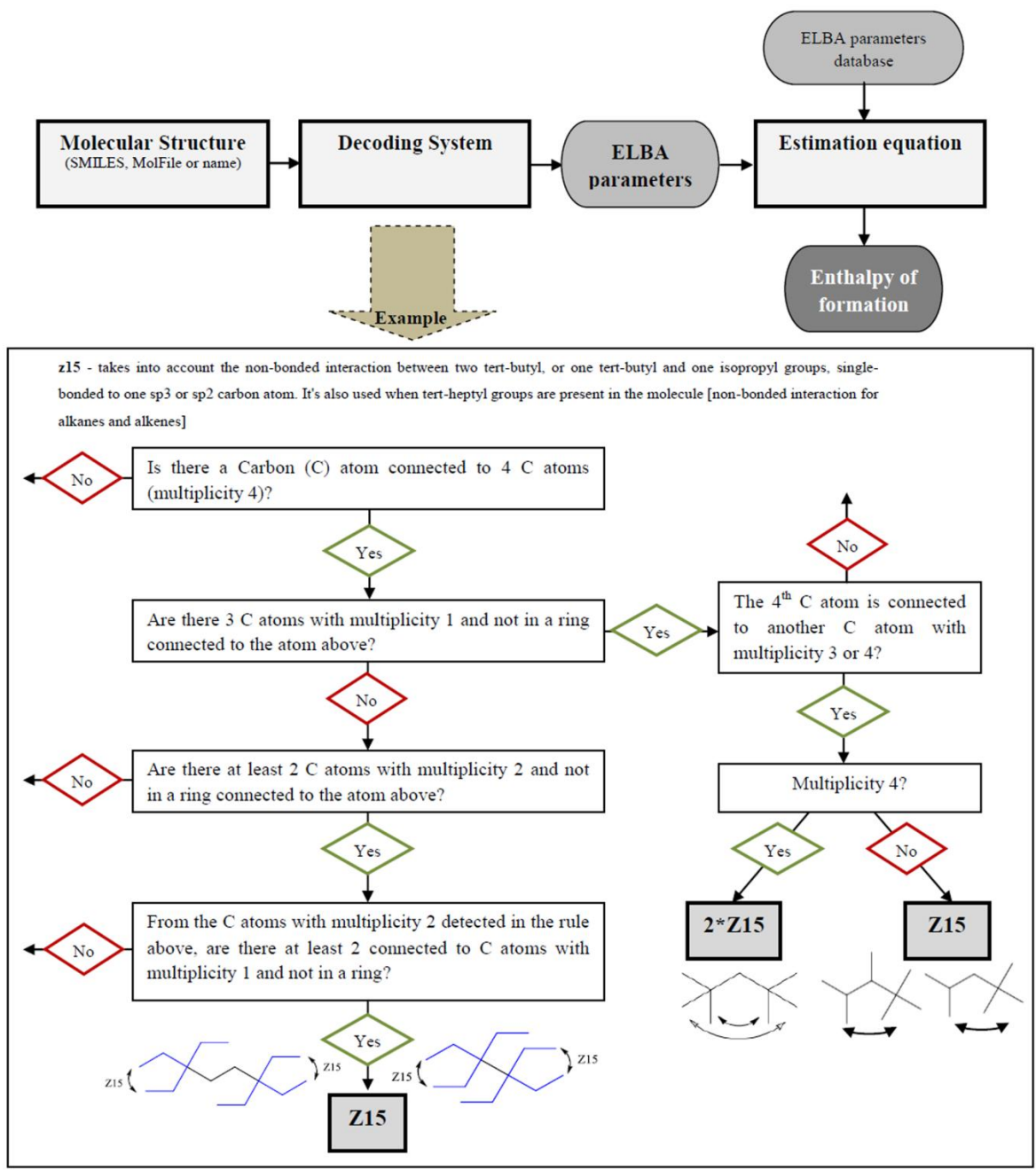

**Figure 1 Diagram representing the implementation of the ELBA method.** The steps needed to calculate the enthalpy of a compound with a specific example of the decoding system for the parameter z15.

### System architecture and requirements

We have designed a relational database to accommodate and organize the heterogeneous data in a way that is compatible with our goals of efficient loading, updating, querying, and eliminating redundancy. The relational model was chosen because it has a good performance and powerful querying of data, using a high-level language, Structured Query Language (SQL) [74]. In addition, it is easy to administer, has wide acceptance, and is well documented, which facilitates the use and development of applications that work on the database [75,76]. To allow a quick, easy, and intuitive access to the data, we have designed a user-friendly web tool based on three main requirements: the visual design, the interaction design, and the functionality. In the development process, we are performing multiple interactions with the following steps: problem analysis and project planning; requirements analysis; system modelling; system implementation; system evaluation; and system maintenance [77-79]. All steps were monitored by users in order to assess the system usefulness, design, and workability [80]. The client/server architecture was adopted to design the system [77]. On the client side, the browser is used for controlling user's input/output and communicating with the web server. The server side is responsible for manipulating the data, communicating with the database and retrieving the data requested from the client side.

## UTILITY AND DISCUSSION

### Data preprocessing and database

The database has been loaded with data collected by the Molecular Energetics Group (Centro de Química e Bioquímica) [76,81]. The database fields can be divided into three categories: structural data, thermochemical data, and the corresponding references (see *Supporting Information*, Table **S1**).

The database consists of twelve tables segmented into five logical categories of data: structural, thermochemical, references, registered users, and database statistics/evolution. The statistics of the database in June, 2012 are displayed in Table 2 and evidence the completeness and representativeness of the data set. Currently, it contains a completely non-redundant set of 2,956 organic compounds, corresponding to more than 15,500 different compound names and synonyms. The analysis of Table 2 shows also that the 2,956 compounds are well characterized in terms of structural properties and are divided into 6 classes, 12 subclasses, and 387 families, according to their constitution and structural arrangement. All the compounds in the current version of the database are characterized with at least one thermochemical property. The database contains 4,687 values for the standard molar enthalpy of formation (crystalline, liquid, and gas phases) and 1,790 values for the standard molar enthalpy of phase change (fusion, vaporization, and sublimation). The property data include the experimental uncertainties [82,83]. The size of ThermInfo dataset is comparable to other widely used databases. The NIST WebBook (2011 version) [49], for instance, contains more than 48,000 organic and inorganic compounds but only about 6,000 have at least a value of enthalpy of formation and/or phase change. The CRC Handbook of Chemistry and Physics (2010 version)[5] contains approximately 10,000 organic compounds, but enthalpies of formation and/or phase change are reported for 1,500.

**Table 2.** ThermInfo database statistics: number of records in the several categories/fields of the dataset (June, 2012).

| Data category | Data field | | Number of records in the database |
|---|---|---|---|
| **Structural Data** | SMILES | | 2,956 |
| | Chemical Structure | | 2,956 |
| | CASRN | | 2,952 |
| | Compound Name | | 2,956 |
| | Synonyms | | 12,598 |
| | Classes | | 6 |
| | Sub-Classes | | 12 |
| | Families | | 387 |
| **Thermochemical Data** | Standard Molar Enthalpy of Formation | Crystalline Phase | 1,461 |
| | | Liquid Phase | 1,486 |
| | | Gas Phase | 1,740 |
| | Standard Molar Enthalpy of Phase Change | Fusion | 83 |
| | | Vaporization | 1,093 |
| | | Sublimation | 614 |
| **Total Number of Compounds** | | | **2,956** |

**Technologies options for the system implementation**

The database is implemented using MySQL [84]. Many of the basic application tools, scripts, and web interfaces, were developed using Hypertext Preprocessor (PHP), a server-side programming language designed especially for the web, with the possibility to be embedded in Hypertext Markup Language (HTML) code. PHP is very well documented, supports intensive transactions, runs fast, and works well with other programming languages chosen for the development of this project, MySQL and JavaScript. Both data presentation and property prediction features allow the user to draw the chemical structure in a JAVA applet (JChemPaint) [85] and export it as a SMILES or 3D MDL MOL file. The use of JavaScript allows a dynamic interaction with the structure. A drawn chemical structure can be converted to a downloadable file format using a Python library (OASA) [86]. Conversions between a given structure identifier and another structure identifier or representation are made by (a not yet completely validated structure-name lookup) the Chemical Identifier Resolver provided by the NCI/CADD group [87], using a simple Uniform Resource Locator (URL) Application Programming Interface (API) scheme or Open Babel [88,89]. Pybel [90] (a Python library that provides access to the Open Babel toolkit) was also used to convert file formats, calculate molecular fingerprint to compare molecules, and to access data and information about structural attributes of the molecule in order to extract the ELBA [60] parameters and predict thermochemical properties.

The web interface is delivered using the open-source Apache web server [91]. Control over access to administrative functions is performed by using Apache Hypertext Access.

**Overview of the ThermInfo web-based system**

A schematic summary of the ThermInfo system architecture is presented in Figure 2, evidencing the relationships between the system main components.

*Data sources*. Data collected by the Molecular Energetics Group[76,81] and data submitted to a temporary database by users, via a web browser (these data are subject to a validation process by an administrator).

*Administrative features*. Enables administrators to interact with the database via a web browser. This interaction involves data validation (manual curation of the data submitted by users, stored in temporary database) and data removal/update.

*Data presentation*. Queries the database and displays information about the compounds in a structured way, via a web browser, based on four types of data-search: quick, advanced, structural, and substructure.

*Properties prediction*. Predicts thermochemical data and displays information in a structured way, via a web browser, based on two types of data input: quick and structural.

ThermInfo is officially available at http://www.therminfo.com or http://therminfo.lasige.di.fc.ul.pt.

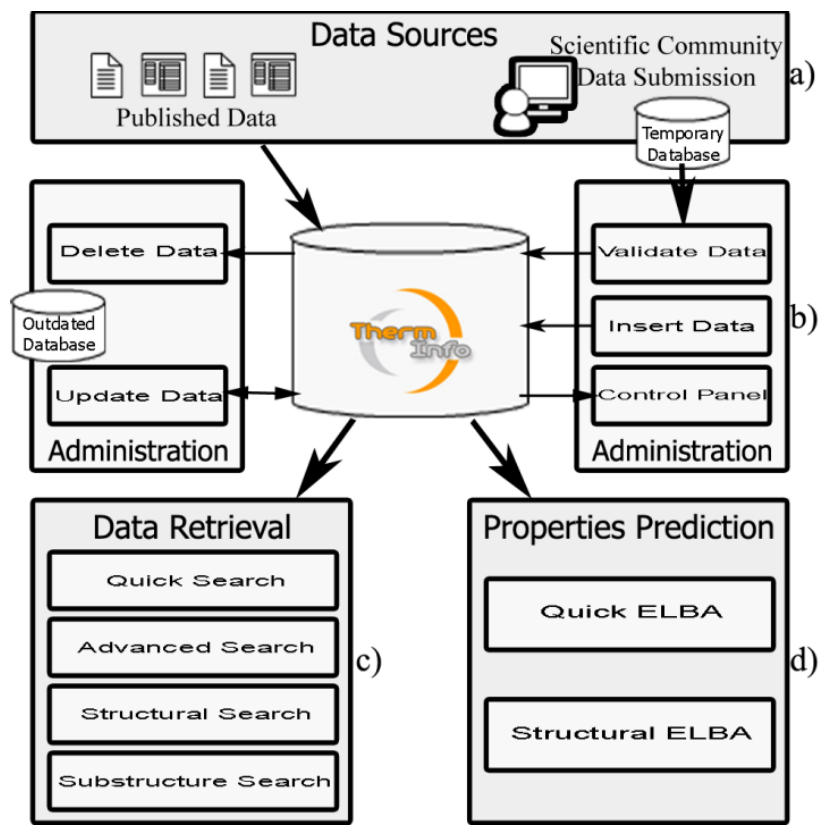

**Figure 2.** Simplified scheme of ThermInfo system architecture. Relationships between the database and the data sources, administrative features, data presentation, and properties prediction. **(a)** data sources; **(b)** administrative features; **(c)** data retrieval; **(d)** properties prediction.

**Dissemination facilities**

ThermInfo has been conceived to provide fast and easy access to information related to structural and thermochemical properties. For this purpose we have implemented a user-friendly web interface to query, insert new data to the database, predict thermochemical properties, and administrate the system. All the features are described in detail in the 'Help' section of the system. In addition, many forms contain guiding messages.

The major features of ThermInfo can be divided into four groups, according to its functions.

*Finding and presenting data*. ThermInfo has multiple search interfaces available, enabling both simple and complex queries (Figure 3). To search a compound in ThermInfo, the user selects the desirable search option and then specifies the query mode or draws a chemical structure. The data are retrieved by the server, which performs a data pre-processing, queries the database, and outputs a results list.

**Figure 3.** Composite screenshot example of data retrieval features. **(1)** The three types of data input: **(a)** Quick Search, term-based search; **(b)** 'Advanced Search', multiple search fields based on specific structural characteristics; **(c)** 'Structural or Substructure Search', based on the molecular structure drawn in a Java applet. **(2)** Search result list: the query description and the list of compounds found in the database. **(3)** Detailed information available for the selected compound.

ThermInfo has four search features available (Table 3): 'Quick Search', which provides a single text box that allows users to search for a chemical compound based on the compound name, the molecular formula, the molecular ID, CASRN, or SMILES; 'Advanced Search', which provides multiple search fields that allow users to limit the search results, based on specific characteristics, namely the compound name, the molecular formula, the physical state, the molecular weight, class, sub-class, family, functional groups, and other; 'Structural Search' and 'Substructure Search', which provide a Java applet that allows users to draw chemical structures or substructures that will be internally converted into a SMILES string (alternatively a SMILES string can be directly typed).

A successful search displays the query description and a summary of the information for the 100 most relevant compounds. By clicking 'View' for a specific compound, the user will be able to obtain all the information available for that compound. The results are ordered according to the selected search type (Table 3).

*User Contributions*. The feature 'Insert Data' is restricted to registered users, to prevent the inclusion of inappropriate data, and allows them to submit data for new compounds. Before disclosing these data, an automatic pre-processing stage and a curation process performed by an administrator are implemented, to verify and validate possibly incorrect data. The main purpose of this feature is to support the expansion of the database by the scientific community. The features 'Suggestion', 'Erroneous Data', 'Question', etc. are within 'Contact Us' and allow an easy interaction between users and the ThermInfo team. This is very important not only to understand the user's questions and needs but also because it allows the participation of the scientific community in the process of assessing data quality.

*System Management*. ThermInfo administrators manage the system via a simple administrative web interface. In 'Administration' new records can be added, updated or deleted. To update or delete records the administrator needs to search for the compound using its molecular ID. If it exists, it will display the compound information that can be deleted or changed. The outdated data are moved to outdated tables. The 'Validate Data' feature allows the administrator to check for new compounds added by users and approve or reject their insertion into the ThermInfo database. The approval or rejection of the new data is automatically reported to the depositor. The 'Control Panel' feature allows the administrator to monitor the current data suppliers, as well as the usage and the growth of the database over time.

**Table 3.** Description of the search types available on each search feature.

| Search Type | Description | Quick | Advanced | Structural | Substructure |
|---|---|:---:|:---:|:---:|:---:|
| Compound Name | The search term is the IUPAC name of a compound. However, searching by using alternative names is also possible for many compounds. Wildcards are not required since the search method retrieves all the compounds whose names contain the search string. The results list will be ordered according to the index of the first occurrence of the query term in the matching string and the difference between the molecular weight of the query structure and the molecular weight of each compound containing the query term, found in the database. When the textual search has no results and knowing the diversity of homophonic names, as well as the propensity for humans to misspell names or misplace letters and numbers, the system retrieves those compounds which the name phonetically matches the search term. The phonetic matching is based on the MySQL's function - soundex(). | ✓ | ✓ | | |
| Molecular Formula | Searches for compounds with the requested chemical formula. In the database, the atoms are in CHXNOS (X = halogen) order, but the search term may be written in any order. The symbol ? can replace the number of atoms of an element. For example, using C?H11 molecular formulas with ? atoms of carbon [? = 2-9] and 11 atoms of hydrogen will be retrieved. | ✓ | ✓ | | |
| Molecular ID | Performs an exact string matching for the unique ID associated with the chemical compounds in the database. Molecular ID is a unique and stable identifier assigned to each compound of the database by ThermInfo. It has the format CONNNNN (N = digit). | ✓ | | | |
| CASRN | Performs an exact string matching for the CAS Registry Number (CASRN) associated with the chemical compounds in the database. It has the format NNNNNNN-NN-N (1-7 digits, hyphen, 2 digits, hyphen, 1 digit). The last digit is a check digit used to verify the validity and uniqueness of the entire number. | ✓ | | | |
| SMILES | This search feature allows to retrieve chemical structures similar to a given query structure, according to a selected similarity threshold. A threshold of 100% (Identical Structures option) retrieves identical chemical structures with different notations (ignoring stereo or isotopic information). Various predefined thresholds between 70-95% are allowed. The calculation of the similarity between the chemical structures is based on Open Babel[88,89] Fingerprints and it is measured using the Tanimoto coefficient [92-96]. The results list will be ordered according to the similarity coefficient (all compounds shown have a similarity score greater than the selected threshold) [71-73] and the difference between the molecular weight of the query structure and the molecular weight of each similar compound found in the database. | ✓ | ✓ | ✓ | ✓ |
| Physical state | Searches for compounds in a certain physical state selected from a drop down menu with the options: gas, liquid or crystal. | | ✓ | | |
| Molecular weight | Searches for compounds matching a specific molecular weight or within a specified interval. | | ✓ | | |
| Class | | | ✓ | | |
| Sub-class | Searches for compounds of a specific class, subclass and/or family, selected from a drop down menu. | | ✓ | | |
| Family | | | ✓ | | |
| Characteristics | Searches for compounds according to the functional groups and other characteristics of the compound. Multiple selections can be made using the check-boxes. | | ✓ | | |

*Properties Prediction*. As mentioned above, it is possible to predict new thermochemical data with ThermInfo. For that purpose, we have developed an ergonomic and functional interface (Figure 4). In the present version of ThermInfo only one estimation method (ELBA) has been implemented. Two different kinds of input can be used. The simplest is 'Quick ELBA', which allows predictions based on a query in a text box (the compound name or its SMILES). If the compound name is used, it will be converted to SMILES using the Chemical Identifier Resolver [87].

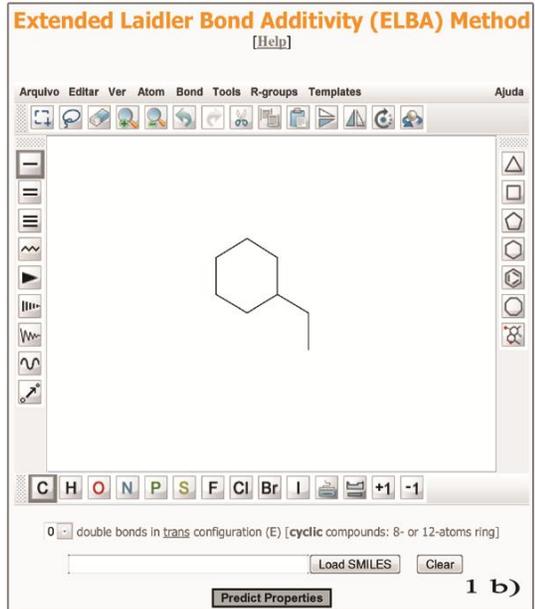

**Figure 4.** Composite screenshot example of properties prediction feature. **(1)** The two types of data input: **(a)** 'Quick ELBA', based on the compound name in a text-box; **(b)** 'Structural ELBA', based on the molecular structure drawn in a Java applet. **(2)** Search result list: the query description and the list of similar compounds found in the database. **(3)** The experimental and predicted values for the properties of the selected compound and the set of ELBA parameters used for the predictions.

The most sophisticated input is under the 'Structural ELBA' feature, which provides a Java applet that allows users to draw the chemical structure of the target compound. This chemical structure will be internally converted by the Java applet into a 3D MDL MOL file to include geometrical information. Once a structure has been drawn or converted from a chemical name into a SMILES string, a structure search is performed in our database. The output includes: the query description (SMILES, compound name, molecular structure, and molecular weight); the option to 'Just Predict Properties' for the target molecule; a list with the target molecule and similar compounds (isomers), if found on the database. By selecting 'Just Predict Properties' for a given compound in the results list, the predicted values and the set of ELBA parameters used for the prediction are presented in tabular form. When 'View' is selected the predicted data are compared to the experimental values. When the 'View' button is not visible, it means that no experimental data are available.

**System performance with respect to response time**

System response time is obviously a very important factor for users but it is also highly dependent on the test environment (for example, internet connection speed, internet traffic level, device characteristics and web browser). This introduces a large amount of error variation into the measured times. ThermInfo performance with respect to response time to retrieve and predict data was assessed in a simple way within a closed Local Area Network environment, to suppress external causes of variance. For that purpose, each compound of a sample of 500 randomly selected compounds was submitted to the system and the time it takes to retrieve and predict properties (excluding the non-hydrocarbon compounds) was measured. This procedure was repeated three times for each search parameter. The average response time and result set size with corresponding standard deviation of the three repetitions of each experiment were recorded (Table 4).

**Table 4.** Recorded average number of hits and response time to retrieve and predict properties with different inputs.

| Input | Average number of hits[a] | std dev[b] | Average response time (seconds)[c] | std dev[b] |
|---|---|---|---|---|
| *Search Performance* | | | | |
| Compound Name | 4.18 | 27.29 | 0.687 | 0.173 |
| Molecular Formula | 9.71 | 13.62 | 0.055 | 0.046 |
| CASRN | 1.00 | - | 0.076 | 0.038 |
| SMILES with similarity threshold = 100% | 3.84 | 8.50 | 1.284 | 0.216 |
| SMILES with similarity threshold = 90% | 4.89 | 9.08 | 1.269 | 0.212 |
| SMILES with similarity threshold = 80% | 11.20 | 23.33 | 1.295 | 0.216 |
| *Prediction Performance* | | | | |
| SMILES | - | - | 3.927 | 4.396 |

[a] Number of compounds found by the searching engine during the search; [b] std dev = standard deviation; [c] Average response time of three repetitions for each experiment.

## CONCLUSIONS AND FUTURE WORK

ThermInfo is a publicly available, web-accessible, information system that brings together critically evaluated values of thermochemical properties of pure substances (for now restricted to organic compounds) and structural data. It also implements a reliable method to predict enthalpies of formation of hydrocarbons. ThermInfo database has already a significant size but it will grow considerably in the near future. This growth will include an extension of the organic compounds database as well as the addition of critically evaluated data for long-lived and transient organic, inorganic, and organometallic molecules in the gas- and condensed-phases. It will also extend the ELBA scheme to other families of compounds, implement a variety of other empirical methods, selected on the basis of their reliability to predict data, and search for new estimation procedures, based on structure-energy relationships and machine learning methods. Finally, ThermInfo scope will also increase in the near future. The database will include values for other molecular properties such as melting point, boiling point, density, refractive index, solubility, standard molar entropy and standard molar Gibbs energy of formation. All this will be possible by implementing a flexible and modular architecture of the database.

ThermInfo strength lies in the data quality, availability (free access), searchability (using different criteria and format input), and, in particular, prediction ability, based on a user-friendly interface that accepts inputs in several formats. It also allows the collaboration of the scientific community, which is encouraged to participate in the project by submitting suggestions and requests for new features and by submitting new data. ThermInfo is available free of charge at [http://www.therminfo.com](http://www.therminfo.com) or [http://www.therminfo.lasige.di.fc.ul.pt](http://www.therminfo.lasige.di.fc.ul.pt).

## Availability and Requirements

*Project name*: ThermInfo
*Project home page*: http://www.therminfo.com
*Operating system*: Platform independent
*System requirements*: Web browser HTML 4.0 compatible with Java (JRE) plug-in and cookies enabled
*Programming language*: PHP, JavaScript, Python
*Database system*: MySQL

## Acknowledgment

ALT and RCS gratefully acknowledge Fundação para a Ciência e a Tecnologia for a doctoral grant (SFRH/BD/64487/2009) and a post-doctoral grant (SFRH/BPD/26610/2006), respectively.

## Supporting Information

A table with *ThermInfo database fields description* was included in Supporting Information.

## Author Contributions

ALT carried out the implementation of the ThermInfo Information System and drafted the manuscript, supervised by AOF and JPL. All authors were involved in the design and testing of the ThermInfo. RCS was responsible for data collection, supervised by JAMS. JPL, RCS, and JAMS developed the ELBA method. All authors read, reviewed, and approved the final version of the manuscript.

# SUPPORTING INFORMATION

**Table S1.** ThermInfo database fields description.

**1. Structural data**, consist of a set of descriptors that specify the molecular structure of the compounds, showing how the atoms are connected, the molecular size, and other properties.

| | |
|---|---|
| **Molecular Identity Descriptor (ID)** | is a unique and stable identifier for the compound, with the format CONNNNN (N = digit). |
| **Compound Name** | is the name provided for a compound, based on the current recommendations of the International Union of Pure and Applied Chemistry (IUPAC). |
| **CAS Registry Number (CASRN)** | is a unique numerical identifier created and assigned to a chemical substance by the Chemical Abstracts Service (CAS). It does not have any chemical significance and it is assigned in sequential order to assure uniqueness. It has the format NNNNNNN-NN-N (1-7 digits, hyphen, 2 digits, hyphen, 1 digit). The right-most digit is a check digit used to verify the validity and uniqueness of the entire number and it is calculated by taking the last digit times 1, the next digit times 2, the next digit times 3 and so on, adding all these up, and computing the sum modulo 10. For example, the CAS number of methanol is 67-56-1: the checksum 1 is calculated as (6 1 + 5 2 + 7 3 + 6 4) = 61; 61 mod 10 = 1.[S1] |
| **Molecular Formula** | identifies each constituent element of a compound by its chemical symbol and indicates the number of atoms of each element in subscript after the chemical symbol. The atoms are in CHXNOS (X = halogen) order. |
| **Chemical Structure** | is a bidimensional structural diagram of the compound in JPG format. |
| **Molecular Weight** | is the mass of one molecule of the compound, relative to the unified atomic mass unit. |
| **Physical State** | are the distinct forms of different phases of matter (gas, liquid or crystalline). |
| **SMILES** | is a specification for describing the structure of chemical molecules using short ASCII strings. This description is case-sensitive. For example, the SMILES for cyclohexane is C1CCCCC1 while for benzene is c1ccccc1.[S2-S4] |
| **USMILES** | is a special and unique SMILES amongst all valid possibilities for a given compound. [S2-S4] |
| **Class, Subclass, Family** | are hierarchical classifications according to the compound structure. |
| **Characteristics** | are tags according to the functional groups present in the molecule and other characteristics of the compound. |

**2. Thermochemical data** are related to the energy released or absorbed in chemical reactions or in physical transformations.[S5,S6]

| | |
|---|---|
| **Standard Molar Enthalpy of Formation** | of a pure substance at 298.15 K is the enthalpy of the reaction where 1 mol of that substance in its standard state is formed from its elements in their standard reference states, all at 298.15 K. The so-called *reference states* of the elements at 298.15 K are their most stable physical states at that conventional temperature. ThermInfo contains values (in kJ mol$^{-1}$) of standard molar |
| | enthalpies of formation at 298.15 K and their associated uncertainties for crystalline, liquid, and gaseous compounds. |
| **Standard Molar Enthalpy of Phase Change** | of a pure substance at 298.15 K is the enthalpy associated with the physical transformation of 1 mol of that substance from one phase to another, where the substance is in its standard state in both phases. Therefore, the standard molar enthalpy of a phase change is simply the difference between the standard molar enthalpies of formation of the substance in the two phases involved. ThermInfo contains values (in kJ mol$^{-1}$) of standard molar enthalpies of fusion (transition from solid to liquid state), vaporization (transition from liquid to gaseous state) and sublimation (transition from solid to gaseous state), at 298.15 K, and their associated uncertainties. |
| **Observations** | provide additional information, if applicable. |

**3. Bibliographic data** provide complete references regarding the source of thermochemical data, including: author(s), journal/book title, year, volume and page(s).

## SUPPORTING BIBLIOGRAPHY